\documentclass[letterpaper, 10 pt, conference]{ieeeconf}  

\IEEEoverridecommandlockouts                              

\usepackage{graphics} 
\usepackage{epsfig} 
\usepackage{times, bm} 
\usepackage{amsmath} 
\usepackage{amssymb} 
\usepackage{hyperref}
\usepackage[font=small,labelfont=bf]{caption}
\usepackage{cite, float}
\usepackage{dblfloatfix}

\usepackage[]{color-edits}
\addauthor{ah}{blue}
\addauthor{ma}{red}

\title{\LARGE \bf
High-Fidelity Accelerated MRI Reconstruction by \\
Scan-Specific Fine-Tuning of Physics-Based Neural Networks
}

\author{Seyed Amir Hossein Hosseini$^{1,2}$,
        Burhaneddin Yaman$^{1,2}$, 
        Steen Moeller$^2$,
        and Mehmet Ak\c{c}akaya$^{1,2}$
        \vspace{-0.4cm}
\thanks{$^{1}$Center for Magnetic Resonance Research and $^{2}$Department of Electrical and Computer Engineering, University of Minnesota, Minneapolis, MN, USA.  {\tt\small e-mails: \{hosse049, yaman013, moell018,  
akcakaya\}@umn.edu}}}

\begin{document}
\maketitle
\thispagestyle{empty}
\pagestyle{empty}

\begin{abstract}
Long scan duration remains a challenge for high-resolution MRI. Deep learning has emerged as a powerful means for accelerated MRI reconstruction by providing data-driven regularizers that are directly learned from data. These data-driven priors typically remain unchanged for future data in the testing phase once they are learned during training. In this study, we propose to use a transfer learning approach to fine-tune these regularizers for new subjects using a self-supervision approach. 
While the proposed approach can compromise the extremely fast reconstruction time of deep learning MRI methods, our results on knee MRI indicate that such adaptation can substantially reduce the remaining artifacts in reconstructed images. In addition, the proposed approach has the potential to reduce the risks of generalization to rare pathological conditions, which may be unavailable in the training data.       
\newline

\end{abstract}

\section{INTRODUCTION}
The field of accelerated MRI reconstruction has substantially benefited from the state-of-the-art advances in deep learning \cite{wang2016accelerating, kwon2017parallel, schlemper2017deep, hammernik2018learning, lee2018deep, akcakaya2019scan, aggarwal2018modl, han2019k, dar2017transfer, hosseini2019dense}. Conventional reconstruction techniques such as parallel imaging \cite{Pruessmann1999, lustig2010spirit} or compressed sensing \cite{lustig2007sparse} typically use fixed and analytical priors for regularized reconstruction. On the contrary, deep learning techniques infer a regularization approach in a data-driven fashion by training neural networks on large databases of data. The trained networks are later used for fast reconstruction of future under-sampled data, usually without any further tuning or modification.  

Even though deep learning has emerged as an effective means for accelerated MRI reconstruction, 
there are a number of remaining challenges.
Primarily, large databases of fully-sampled data may not be available to supervise learning in the training phase. Some studies have addressed this issue by using transferred learning from available large datasets \cite{han2018deep, dar2017transfer}. In this approach, a network is pre-trained on another large dataset, which can also include non-medical images, and then re-trained on a smaller dataset of the specific imaging application in a supervised manner. 
Additionally, both pre-training and transfer datasets may lack examples of rare and/or subtle pathologies, increasing the risks of failure in generalization to such situations \cite{eldar2017challenges, knoll2020advancing}.

In this study, we propose to fine-tune a pre-trained neural network for MRI reconstruction in a scan-specific manner using transfer learning. 
The main challenge in our proposed method is the necessity of tuning the pre-trained reconstruction network for the data to be reconstructed, which itself is undersampled. This also forms the main difference to conventional transfer learning that re-trains on smaller fully-sampled datasets \cite{dar2017transfer,han2018deep}. To tackle the lack of reference data in this scenario, we propose to use a recently developed self-supervised training method called self-supervised learning via data under-sampling (SSDU) to fine-tune the networks in a scan-specific manner \cite{yaman2019self, yaman2019selfMRM}.
The proposed approach is tested on a knee MRI dataset, and compared with the scenario where the undersampled data is directly reconstructed using a database-trained network without any further scan-specific fine-tuning.  

\section{MATERIALS AND METHODS}
\subsection{Problem Formulation}
Let $\mathbf{y}_\Omega$ be the undersampled noisy data from a multi-coil MRI system, where $\Omega$ denotes the under-sampling pattern, 
and ${\bf x}$ be the corresponding underlying image. The The forward model for this system is given as:
\begin{equation}
\label{equ:forwardmodel}
\mathbf{y}_\Omega = \mathbf{E}_\Omega\mathbf{x} + \mathbf{n}, 
\end{equation}
where ${\bf E}_\Omega: {\mathbb C}^{M \times N} \to {\mathbb C}^P$ is the forward encoding operator that includes a partial Fourier matrix specified by the k-space locations in $\Omega$ and the sensitivities of the receiver coil array \cite{Pruessmann1999}, and ${\bf n} \in {\mathbb C}^P$ is the measurement noise. The inverse problem pertinent to Equation (\ref{equ:forwardmodel}) is typically solved using a regularized least squares approach:
\begin{equation}\label{equ:objecivefunction}
\arg \min_{\bf x} \|\mathbf{y}_\Omega-\mathbf{E}_\Omega\mathbf{x}\|^2_2 + {\mathcal R}(\mathbf{x}),
\end{equation}
where the first term enforces consistency with measurement data and $\mathcal{R}(\cdot)$ denotes a regularizer. 
Recently deep learning has been used to learn data-driven priors for the regularization term using algorithm unrolling \cite{gregor2010learning}.

\subsection{Unrolled Network Database Training}
\begin{figure} [!t]
     \begin{center}
             \includegraphics[width=\columnwidth]{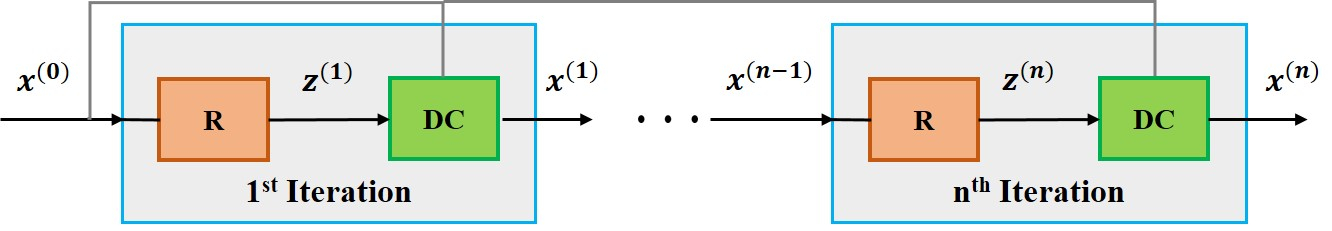}
     \end{center}
     \vspace{-.3cm}
    \caption 
    {\label{fig:netarchi} The unrolled neural network architecture for $n$ iterations, each consisting of a CNN unit to proxy regularization (R) and a data consistency (DC) unit to ensure data fidelity.}
    \vspace{-.2cm}
\end{figure}

A variable splitting approach with quadratic relaxation 
\cite{fessler2020optimization} can be used to solve the optimization problem in (\ref{equ:forwardmodel}) by defining an auxiliary variable $\mathbf{z}$, as follows:
\begin{align}\label{Eq:quardatic-relaxation}
\arg \min_{\bf x,z} \|\mathbf{y}_\Omega-\mathbf{E}_\Omega\mathbf{x}\|^2_2 +\beta \lVert\mathbf{x}-\mathbf{z}\rVert_{2}^2 +\cal{R}(\mathbf{z}),
\end{align}
where $\beta$ is the parameter of the quadratic penalty. 
This objective function can be solved via alternating minimization: 
\begin{subequations}
\vspace{-.3cm}
\begin{align}
& \mathbf{z}^{(i)} = \arg \min_{\bf z}\beta \lVert\mathbf{x}^{(i-1)}-\mathbf{z}\rVert_{2}^2 +\cal{R}(\mathbf{z})\label{Eq:recons3a}
\\
& \mathbf{x}^{(i)} = \arg \min_{\bf x}\|\mathbf{y}_\Omega-\mathbf{E}_\Omega\mathbf{x}\|^2_2 +\beta\lVert\mathbf{x}-\mathbf{z}^{(i)}\rVert_{2}^2,\label{Eq:recons3b}
\end{align}
\end{subequations}
where $i$ denotes the iteration number. Physics-based DL MRI methods unroll this iterative process for a fixed number of iterations. Each unrolled iteration consists of two units; a trainable unit with CNNs to proxy the regularization update at sub-problem (\ref{Eq:recons3a}) and a linear unit to enforce data consistency by solving sub-problem (\ref{Eq:recons3b}) (\textbf{Figure \ref{fig:netarchi}}). While the latter has a closed-form solution, a conjugate gradient (CG) method is typically used to avoid the large matrix inversion in multi-coil MRI parallel imaging \cite{aggarwal2018modl}.
The full unrolled neural network is then trained end-to-end in a supervised manner to learn the CNN parameters in a data-driven manner by minimizing 
\begin{equation}
    \min_{\bm \theta} \frac1N \sum_{i=1}^{N} \mathcal{L}\Big({\bf x}^i_{ref}, \: f({\bf y}_\Omega^i, {\bf E}_\Omega^i; {\bm \theta}) \Big),
\end{equation}
where ${\bf x}^i_{ref}$, ${\bf y}_\Omega^i$ and ${\bf E}_\Omega^i$ are the reference fully-sampled image, under-sampled k-space data and forward encoding operator for slice $i$ respectively, and $N$ is the number of training slices. ${\bm \theta}$ represents the trainable parameters of the network and $f({\bf y}_\Omega^i, {\bf E}_\Omega^i; {\bm \theta})$ denotes the network output for slice $i$ with the corresponding measurements and forward encoding operator. $\mathcal{L}(\cdot, \cdot)$ is a loss function that measures the dissimilarity between the network output and reference fully-sampled image, e.g. $\ell_1$, $\ell_2$ or $\ell_1$-$\ell_2$ loss \cite{knoll2020deep, yaman2019selfMRM}.

\subsection{Proposed Scan-Specific Fine-Tuning of the Network}
Once the unrolled neural network is trained on a database of paired under-sampled and fully-sampled k-space data, it can be readily used to reconstruct future under-sampled data via a feed-forward propagation through the network. 
Note that the regularization CNNs remain fixed regardless of future data specifics or changes, or in the face of database biases with respect to rare pathologies. We hypothesize that scan-specific fine-tuning of the network parameter using only the under-sampled k-space data of interest may further improve the reconstruction performance.

Since fully-sampled reference data is not available during the fine-tuning phase, we propose to use a recently developed self-supervised training method, called SSDU for fine-tuning the network \cite{yaman2019self, yaman2019selfMRM}. SSDU partitions the measurement data $\mathbf{y}_\Omega$ into two sets $\mathbf{y}_\Theta$ and $\mathbf{y}_\Lambda$, where the former is used in the data consistency unit of the network (Equation \ref{Eq:recons3b}) during training and the latter is used to define a loss function in k-space.
Following the SSDU approach, we propose the following loss function for the fine-tuning phase:
\begin{equation}
    \min_{\bm \theta} \mathcal{L}\Big({\bf y}_{\Lambda}, \: {\bf E}_{\Lambda} \big(f({\bf y}_{\Theta}, {\bf E}_{\Theta}; {\bm \theta}) \big) \Big),
\end{equation}
where ${\bf E}_{\Lambda}$ transforms the network output image into coil k-space domain, so the loss can be defined with respect to the k-space points $\mathbf{y}_\Lambda$. We note that the network parameters $\theta$ are initialized with the database-trained network values. These parameters are then fine-tuned
, only using the same data to be reconstructed. Thus, ${\bf y}_{\Theta}$ is used as the data input to the neural network, whose parameters are tuned to best estimate ${\bf y}_{\Lambda}$ at the output based on the loss function. During the final reconstruction, the complete set of measurement data $\mathbf{y}_\Omega$ is then fed into the finely-tuned network.  

\subsection{Knee MRI Datasets} \label{sec:3a}
Fully-sampled knee MRI datasets were obtained from the New York University (NYU) fastMRI initiative database \cite{fastmriRadiologyAI}. Data were acquired on a clinical $3$T system (Magnetom Skyra, Siemens, Germany) with a 15-channel knee coil, using 2D turbo spin-echo sequences in coronal orientation with proton-density (Coronal PD) and proton-density with fat suppression (Coronal PD-FS) weightings. Relevant imaging parameters were: resolution = $0.49 \times 0.44$mm$^2$, slice thickness = $3$mm , matrix size = $320\times368$ for both datasets.
Uniform equi-spaced undersampling with an acceleration rate of4 and with 24 ACS lines were used \cite{fastmriRadiologyAI}. Coil sensitivity maps were generated by ESPIRiT using a $24\times24$ central window for each slice \cite{uecker2014espirit}.

The central $300$ slices from $15$ subjects were used for each sequence during the supervised training phase. Testing was performed on slices from different subjects, both with and without scan-specific fine-tuning of the networks. For the scan-specific fine-tuning phase, the uniformly under-sampled data were split into two sets. To this end, 40\% of the measurement data were randomly selected based on a Gaussian distribution to define the loss points in k-space, as optimized in \cite{yaman2019selfMRM, yaman2019self}, while the remainder of the measurements were used during training for data consistency. 

\begin{figure*} [!t]
    \begin{center}
            \includegraphics[width=\textwidth]{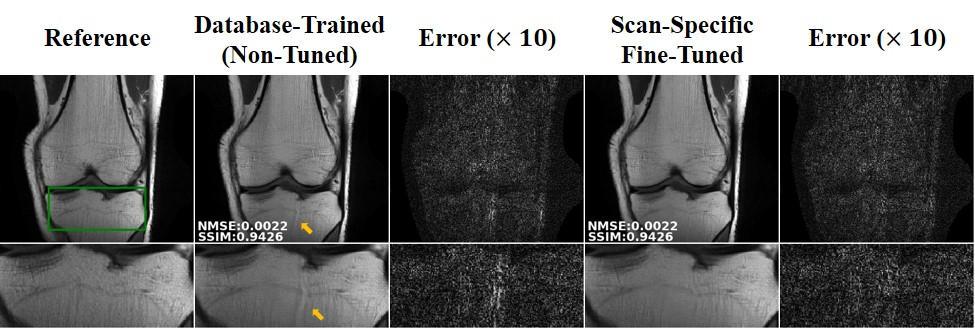}
    \end{center}
     \vspace{-.3cm}
    \caption 
    {\label{fig:coronalpd} A representative slice from the Coronal PD dataset, uniformly under-sampled at an acceleration rate of $4$ and reconstructed with a database-trained neural network (Non-Tuned) and the proposed scan-specific fine-tuned version of the same network (Tuned). Error images with respect to the reference fully-sampled image, scaled by a factor of 10, are provided for ease of comparison. The green rectangle in the reference image mark the zoom-up area shown in the second row for each method. Scan-specific fine tuning of the trained network removes the residual aliasing artifacts which are visible in the original reconstructed image (pointed to by arrows). Quantitative metrics, shown on the lower-left corner of the images, remain unchanged.}
    \vspace{-.2cm}
\end{figure*}

\begin{figure*} [!b]
    \begin{center}
            \includegraphics[width=\textwidth]{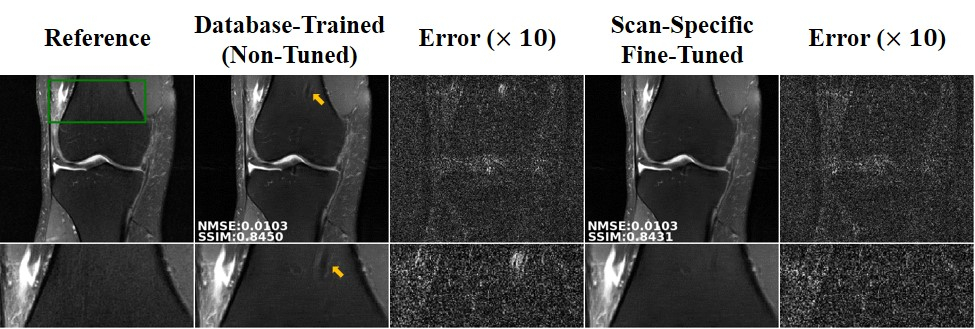}
    \end{center}
     \vspace{-.3cm}
    \caption 
    {\label{fig:coronalpdfs} A representative slice from the Coronal PD-FS dataset, uniformly under-sampled at an acceleration rate of $4$ and reconstructed with a database-trained neural network (Non-Tuned) and the proposed scan-specific fine-tuned version of the same network (Tuned), as well as error images scaled by a factor of 10. The green rectangle in the reference image mark the zoom-up area shown in the second row for each method. The proposed scan-specific fine-tuning approach visibly removes the residual artifacts (pointed to by arrows) in the non-tuned image, while the quantitative metrics fail to reflect this improvement.}
    \vspace{-.2cm}
\end{figure*}

\subsection{Implementation Details}
The network in {\bf Fig. \ref{fig:netarchi}} was unrolled for $10$ iterations.  
DC units implement CG with $10$ internal iterations, while regularization units share a residual network (ResNet) \cite{timofte2017ntire}, consisting of $15$ sub-residual blocks each having two convolutional layers of kernel size = $3\times3$ and output channels = $64$. A ReLU activation function is applied after the first convolutional layer in each sub-residual block whose output is scaled by a factor of $0.1$. Two input and output convolutional layers without any nonlinear activation function are to match number of desired channels, in addition to the sub-residual blocks. Regularization units share parameters across unrolled iterations \cite{aggarwal2018modl}, leading to a total of 592,129 including the quadratic relaxation parameter. Adam optimizer was used to minimize a normalized $\ell_1-\ell_2$ loss \cite{knoll2019deep, yaman2019self, yaman2019selfMRM} 
during both supervised training and scan-specific fine-tuning of the network. 
Training was performed over $100$ and $15$ epochs using a learning rate of $10^{-3}$ and $10^{-4}$, for the database and scan-specific training phases, respectively.


\section{RESULTS}
{\bf Fig. \ref{fig:coronalpd}} depicts a representative slice from the coronal PD dataset, reconstructed with the database-trained neural network and the proposed scan-specific fine tuning of the same network. Error images with respect to the reference fully-sampled image are also displayed after scaling by a factor of $10$ for ease of comparison. While most of aliasing artifacts have disappeared in the database-trained reconstruction, some residual artifacts are still observed. The proposed scan-specific fine-tuning approach successfully removes these residual aliasing artifacts that are visible in the non-tuned image. Quantitative metrics do not change between the reconstructed images.  

{\bf Fig. \ref{fig:coronalpdfs}} shows a representative slice from the coronal PD-FS dataset, reconstructed with the database-trained network and also with the same network architecture after its scan-specific fine-tuning. Error images with respect to the reference, scaled by a factor of $10$ are also displayed. Similar to the previous case, the scan-specific fine-tuning approach reduces the remaining artifacts that are visible in the non-tuned images. Quantitative evaluation fails to reflect this improvement. 

\section{DISCUSSION}
In this study, we proposed a scan-specific fine-tuning method for improving the quality of physics-driven deep learning MRI reconstruction. While the proposed approach benefits from data-driven regularization feature of deep learning methods, it also adapts to data specifics by fine-tuning a trained network parameters for the under-sampled data to be reconstructed. Therefore, the proposed method can reduce the risks of incomplete generalization to future data which may particularly be a concern for medical imaging applications \cite{eldar2017challenges, knoll2020advancing}.

In contrast to previous scan-specific accelerated MRI reconstruction techniques, which train neural networks on limited calibration data for parallel imaging albeit without an explicit regularizer \cite{akcakaya2019scan, hosseini2019sraki, hosseini2019accelerated, hosseini2019acceleratedArxiv, zhang2019optimized, Zhang2019Asilomar, kim2019loraki}, the proposed approach transfers learning from pre-trained networks. Thus, larger neural networks with higher capacities can be trained, further improving the reconstruction performance \cite{knoll2020deep}.
In addition, the fine-tuning phase is not as time-consuming, since only one dataset is used and full convergence can be achieved quickly ($\sim15$ seconds, in this study). 
Nonetheless, while the extra fine-tuning phase increases the total reconstruction time, the results indicate that the proposed method can considerably reduce the remaining aliasing artifacts of reconstruction performed with a database-trained network. In addition to offering improved reconstruction quality, potential database-training biases against rare or subtle pathologies may be avoided in this way \cite{eldar2017challenges, knoll2020advancing}, which warrants further investigation. Finally, the current study is limited to a $4$-fold acceleration rate. A future study assessing the scan-specific fine-tuning at different acceleration rates is warranted.

\section{CONCLUSION}
We proposed a scan-specific transfer learning method that uses self-supervised training to adapt a physics-driven reconstruction network trained on a large database to the dataset of interest for high-fidelity accelerated MRI reconstruction. 

\section*{ACKNOWLEDGMENTS}
This work was supported by NIH P41EB027061, U01EB025144; NSF CAREER CCF-1651825. Knee MRI data were obtained from the NYU fastMRI initiative database \cite{fastmriRadiologyAI}. NYU fastMRI database was acquired with the relevant institutional review board approvals as detailed in \cite{fastmriRadiologyAI}. NYU fastMRI investigators provided data but did not participate in analysis or writing of this report. A listing of NYU fastMRI investigators, subject to updates, can be found at \url{fastmri.med.nyu.edu}.

\bibliographystyle{IEEEbib}
\bibliography{reference}

\end{document}